\input{aipcheck}

\documentclass[
    ,final            
  ]
  {aipproc}

\layoutstyle{8x11single}

\begin{document}

\title{Measurement of ortho-Positronium Properties in Liquid Scintillators}


\author{S. Perasso}{
  address={Laboratoire AstroParticule et Cosmologie, Paris}
}

\author{G. Consolati}{
  address={Department of Aerospace Science and Technology, Politecnico di Milano}
}

\author{D. Franco}{
  address={Laboratoire AstroParticule et Cosmologie, Paris}
}

\author{S. Hans}{
  address={Brookhaven National Laboratory}
}

\author{C. Jollet}{
  address={IPHC, Universit\'e de Strasbourg}
}

\author{A. Meregaglia}{
  address={IPHC, Universit\'e de Strasbourg}
}

\author{A. Tonazzo}{
  address={Laboratoire AstroParticule et Cosmologie, Paris}
}

\author{M. Yeh}{
  address={Brookhaven National Laboratory}
}

\begin{abstract}
Pulse shape discrimination in liquid scintillator detectors is a well-established technique for the discrimination of heavy particles from light particles.
Nonetheless, it is not efficient in the separation of electrons and positrons, as they give rise to indistinguishable scintillator responses.
This inefficiency can be overtaken through the exploitation of the formation of ortho-Positronium (o-Ps), which alters the time profile of light pulses induced by positrons.

We characterized the o-Ps properties in the most commonly used liquid scintillators, i.e. PC, PXE, LAB, OIL and PC~+~PPO.
In addition, we studied the effects of scintillator doping on the o-Ps properties for dopants currently used in neutrino experiments, Gd and Nd.
Further measurements for Li-loaded and Tl-loaded liquid scintillators are foreseen.
We found that the o-Ps properties are suitable for enhancing the electron-positron discrimination.

\end{abstract}

\maketitle


\section{Introduction}
Pulse shape discrimination (PSD) is a fundamental technique for background rejection in scintillator detectors.
It relies on the different time profiles of the emitted scintillator light to solicitations from particles with different energy losses.
For this reason, it is particularly effective for discriminating heavy particles (ions, alphas or protons) from light particles (electrons, positrons).
Even so, electrons and positrons generate similar light pulses and therefore PSD fails in discriminating them.

PSD discrimination between electrons and positrons can be recovered exploiting the deformation of the positron induced light pulse shape caused by the formation of positronium, a meta-stable positron-electron bound state.
Depending on the system total spin, the positronium can be formed in the singlet state, called para-Positronium (p-Ps, B.R. 25$\%$), or in the triplet state, called ortho-Positronium (o-Ps, B.R. 75$\%$).
Due to the charge-conjugation invariance, p-Ps decays into two 511 keV gammas, with a mean life in vacuum of 125 ps, where o-Ps decays into three gammas of total energy equal to 2$m_e$, with a mean life in vacuum of 142 ns.
Nevertheless, in matter the o-Ps mean life is strongly reduced by the interactions with the surrounding medium.
Chemical reactions (oxidation, compound compositions), magnetic effects (spin flip) or pick-off (positron annihilation with an anti-parallel spin electron of the medium) cause the o-Ps to decay into two gammas with a mean life of a few ns.
The decay into three gammas is usually reduced to a negligible fraction.

An o-Ps mean life of a few ns is not enough for separating the positron energy deposition from the annihilation gammas in scintillator detectors, but it can still induce a positron pulse shape deformation strong enough to enhance electron-positron discrimination.
This enhancement could be exploited in reactor neutrino experiments \cite{DC:2012, DB:2012, Reno:2012}, in which
anti-neutrino detection is performed via inverse beta decay
\begin{equation}
  \bar{ \nu_e } + p \rightarrow e^+ + n.
\end{equation}
The positron energy deposition and its following annihilation provide the prompt event and the neutron absorption the delayed event. 
An o-Ps enhanced PSD could allow to suppress the accidental background and to reject the residual correlated background, given by the cosmogenic $\beta^--$n emitters $^9$Li and $^8$He.

The same technique can be also exploited for background rejection in underground low background experiments, like Borexino \cite{Borexino:2009} and SNO+ \cite{SNO:2005}, in which several $\beta^+$ emitters ($^8$B, $^9$Be, $^{10}$C) produced by cosmic muon spallation, constitute an unavoidable source of background.
It has already been successfully applied in Borexino for the detection of solar \textit{pep} neutrinos, where the rejection of cosmogenic $^{11}$C is essential for the signal extraction.

The characterization of o-Ps properties, i.e. the measurement of its formation probability and lifetime, 
has been performed for the most commonly used liquid scintillators, namely PC, LAB, PXE and OIL, and for the mixture PC + PPO \cite{Franco:2010rs}.
Furthermore, the effect of scintillator doping on the o-Ps properties has been studied for different dopants: Gd, commonly used in reactor neutrino experiments to enhance the neutrino signal identification, and Nd, originally considered as neutrinoless double beta decay signal source in the SNO+ experiment.
Forthcoming measures will characterize also Li-loaded and Tl-loaded liquid scintillators, the latter being chosen as new dopant for the SNO+ scintillator.

\section{Experimental Apparatus}
The o-Ps properties (lifetime and formation probability) in scintillator are measured by means of a 0.8 MBq $^{22}$Na source in a fast coincidence system.
The radio-isotope $^{22}$Na decays $\beta^+$ to $^{22}$Ne with a B.R. of 89.2$\%$ (Q-value = 2.842 MeV).
In 99.944$\%$ of cases, the decay is to an excited state of $^{22}$Ne, which de-excites emitting a 1.274 MeV gamma.
The o-Ps properties are extrapolated from the measurement of the time distance between the de-excitation gamma (prompt signal) and the positron annihilation gammas (dealyed signal).
  
A droplet of inert solvent containing the source is deposited between two identical supports, 1 cm in radius, each composed by two 7.5 $\mu$m thick Kapton foils.
Kapton is an aromatic polyimide material where positrons are known not to form Positronium \cite{McGuire:2006, Plotlowski}.
The source sandwich is placed into a vial containing the scintillator sample.

The vial is in turn positioned between two plastic scintillator detectors, each coupled to a PMT.
The threshold on the first detector is set to 900 keV to isolate the $^{22}$Ne de-excitation gamma, while the second detector selects gammas between 350 keV and 500 keV, in order to accept the annihilation gammas but reject the de-excitation gamma backscattering.

\begin{figure}
  \includegraphics[height=.25\textheight]{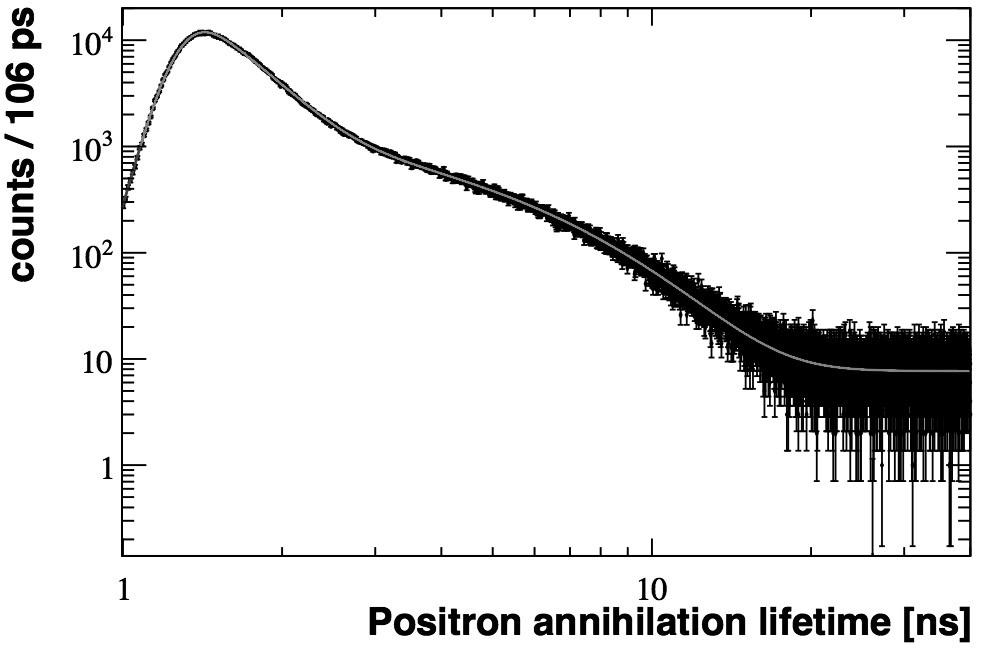}
  \caption{Positron annihilation time spectrum for the PXE sample, with the fitted function superimposed.}
  \label{fig:fit}
\end{figure}

A constant fraction discriminator generates a fast timing signal at each trigger from the two detectors.
A time to amplitude converter generates a signal linearly increasing with time, starting at the prompt signal and stopping at the delayed signal.
The output is digitized by a 4096 channels ADC.
Figure \ref{fig:fit} shows an example of resulting time spectrum.

The system is calibrated using a $^{60}$Co source, and each ADC channel is found to corresponds to 10.6 ps.
The resolution is measured with the fast coincidence (182 ps) gammas from $^{207}$Bi and found to be 0.28 ns at FWHM.

\section{Characterization of o-Ps in Liquid Scintillators}
Positronium has been characterized in common liquid scintillators (PC, PXE, LAB, OIL, PC + PPO) at room temperature.
Each sample has been flushed with Nitrogen and then measured with a $\sim 1 \times 10^6$ event statistics, repeatedly from 3 to 5 times, in order to account for possible variations in the experimental conditions (temperature, Oxygen concentration).

The obtained time distributions are fitted using the RooFit toolkit \cite{Roofit}, embedded in the ROOT package and based on MINUIT.
The fit function is
\begin{equation}
F(t) = \chi ( t > t_0 ) \cdot \left ( \sum_{k=1,2} \frac{A_k}{\tau_k} \cdot e^{-t/\tau_k} + C \right )
\label{func:fit1}
\end{equation}
where $\chi$ is the step function and A$_i$ and $\tau_i$ are the amplitude and mean life of the \textit{i}-th component:
the first accounts for the effective contribution of direct annihilation and p-Ps, while the second accounts for the o-Ps.
The constant term accounts for accidentals.
The function $F(t)$ is convoluted with the detector resolution function, the sum of two gaussians with the same mean value, but different sigmas ($\sigma_1$ = 110 ps and $\sigma_2$ = 160 ps)  and relative weights (g$_1$ = 0.8 and g$_2$ = 0.2).
The obtained normalized $\chi^2$ are in the range between 0.85 and 0.94 and the statistical errors are at the per mil level.

The o-Ps mean life is given directly by the fit parameter $\tau_2$, while the formation probability $f_2$ is computed as the fraction of o-Ps relative to the annihilations in the scintillator
\begin{equation}
f_2 = \frac{A_2}{A_1 + A_2 - A_K}.
\end{equation}
The fraction of direct annihilations in Kapton, A$_K$, has been extrapolated from dedicated measurements with different number of Kapton layers and estimated to be
(20.6 $\pm$ 0.2)$\%$.

The systematic uncertainty is evaluated 
from the dispersion of the measured mean lives from the weighted average of the corresponding scintillator sample.
The distribution of all the dispersions is fitted to a gaussian, whose standard deviation is taken as the systematic error.
The same procedure is followed for the formation probability.
The resulting systematic errors are 0.03 ns and 0.5$\%$.

The final results are summarized in Table \ref{tab:liqscintres}.
All the scintillators exhibit an o-Ps formation probability around 50$\%$ and a mean life of $\sim$3 ns.
\begin{table}
\begin{tabular}{lrr}
\hline
     \tablehead{1}{r}{b}{Scintillator}
  & \tablehead{1}{r}{b}{o-Ps fraction}
  & \tablehead{1}{r}{b}{lifetime [ns]} \\
\hline
PC				&      0.485 $\pm$ 0.005      &      2.96 $\pm$ 0.03 \\
PXE				&      0.466 $\pm$ 0.005      &      2.74 $\pm$ 0.03 \\    
LAB				&      0.542 $\pm$ 0.005      &      3.08 $\pm$ 0.03 \\     
OIL				&      0.506 $\pm$ 0.005      &      3.04 $\pm$ 0.03 \\     
PC + 1.5 gl PPO	&      0.512 $\pm$ 0.005      &      3.12 $\pm$ 0.03 \\
\hline
\end{tabular}
\caption{Measured values of o-Ps formation probability and lifetime for the different samples of liquid scintillators.}
\label{tab:liqscintres}
\end{table}

\section{Characterization of doped Liquid Scintillators}
The same instrumental setup and analysis tool have been used to characterize the o-Ps in doped liquid scintillators.
The dopants tested so far are Gd, a neutron capture enhancing isotope, and Nd, originally considered for SNO+.
Both dopants have been tested at different concentrations between 0.01$\%$ and 0.5$\%$ in the common solvent LAB.
Further measurements are foreseen with Tl-loaded and Li-loaded liquid scintillator.

The positron annihilation time spectrum is fitted with the superposition of three exponentials and a constant:
the first two exponentials describe the direct annihilation and the p-Ps components, the third one is an effective contribution from two and three gamma o-Ps decays and the
consent accounts for accidentals.
The detector resolution is modeled with a single gaussian with $\sigma$ = 120 ps.

The lifetime of o-Ps decay into two gammas, $\tau_2$, can be extracted from the fitted parameter $\tau$ according to the relation $\tau_2$ = $\tau_3\tau/(\tau_3 - \tau)$, $\tau_3$ being the o-Ps lifetime in vacuum (142 ns).
Then, the fraction of decays into three gammas is $f_3$ = $\tau/\tau_3$ and the one of decays into two gammas is $f_2$ = 1 - $f_3$.
The o-Ps probability formation, $f$, is computed taking into account the different detection efficiencies for the two gamma channel ($\epsilon_2$) and the three gamma channel ($\epsilon_3$) and is given by
\begin{equation}
f = \frac{A_3\tau_3}{(A_A + A_3 - A_K) \cdot \tau_3 + (A_A - A_K)(\epsilon_3/\epsilon_2 - 1)\tau}
\label{equ:f}
\end{equation}
where A$_A$ = A$_1$ + A$_2$ and A$_K$ is the fraction of positron annihilations in Kapton.
Since the two efficiencies could not be precisely measured, $f$ is computed for the two extreme cases $\epsilon_3$ = 0 and $\epsilon_3$ = $\epsilon_2$.
The average value is taken as measure of $f$, while the semi-difference is added to the systematics.
The other contribution to systematics comes from the deviation of measurements from the weighted averages, described in the previous section.
Finally, the statistical uncertainty on $f$ resulting from the error propagation in Eq. (\ref{equ:f}) is taken into account.
The systematic uncertainties on lifetime is evaluated accordingly, with an additional contribution coming from the setup calibration.
To summarize, the relative uncertainty on the o-Ps formation probability is 1.9$\%$, while the one on the mean life is $\sim$1$\%$.

\begin{table}
\begin{tabular}{ccc}
\hline
     \tablehead{1}{r}{b}{Gd concentration}
  & \tablehead{1}{r}{b}{o-Ps fraction}
  & \tablehead{1}{r}{b}{lifetime [ns]} \\
\hline
0		   &      0.544 $\pm$ 0.008      &      3.05 $\pm$ 0.03 \\
0.01$\%$ 	   &      0.554 $\pm$ 0.008      &      3.07 $\pm$ 0.03 \\    
0.05$\%$     &      0.540 $\pm$ 0.008      &      3.05 $\pm$ 0.03 \\     
0.08$\%$     &      0.537 $\pm$ 0.008      &      3.04 $\pm$ 0.03 \\     
0.1$\%$       &      0.529 $\pm$ 0.008      &      3.09 $\pm$ 0.03 \\
0.45$\%$     &      0.406 $\pm$ 0.008      &      3.02 $\pm$ 0.03 \\
\hline
     \tablehead{1}{r}{b}{Nd concentration}
  & \tablehead{1}{r}{b}{o-Ps fraction}
  & \tablehead{1}{r}{b}{lifetime [ns]} \\
\hline
0		   &      0.537 $\pm$ 0.013      &      3.15 $\pm$ 0.04 \\
0.05$\%$     &      0.527 $\pm$ 0.013      &      3.11 $\pm$ 0.04 \\     
0.1$\%$       &      0.494 $\pm$ 0.013      &      3.17 $\pm$ 0.04 \\     
0.3$\%$       &      0.460 $\pm$ 0.013      &      3.15 $\pm$ 0.04 \\
0.5$\%$       &      0.402 $\pm$ 0.013      &      3.15 $\pm$ 0.04 \\
\hline
\end{tabular}
\caption{Measured values of o-Ps formation probability and lifetime for the Gd-loaded and Nd-loaded liquid scintillators at different concentrations.}
\label{tab:a}
\end{table}
Table \ref{tab:a} reports the results for the Gd-loaded and Nd-loaded liquid scintillators at the different concentrations.
Both of the dopants show the same effect on the o-Ps properties: the mean life remains stable under dopant concentration variations, while the formation probability decreases from $\sim$50$\%$ at concentrations of the order of few tenths per mil to $\sim$40$\%$ at concentrations around 0.5$\%$.

\begin{theacknowledgments}
We acknowledge the financial support from the ANR NuToPs project (ANR-11-JCJC-SIMI4) and from the UnivEarthS Labex program of Sorbonne Paris Cit\'e (ANR-10-LABX-0023 and ANR-11-IDEX-0005-02).
\end{theacknowledgments}





\bibliographystyle{aipproc}   


\IfFileExists{\jobname.bbl}{}
 {\typeout{}
  \typeout{******************************************}
  \typeout{** Please run "bibtex \jobname" to optain}
  \typeout{** the bibliography and then re-run LaTeX}
  \typeout{** twice to fix the references!}
  \typeout{******************************************}
  \typeout{}
 }


\end{document}